\documentclass[preprint,12pt]{elsarticle}

\usepackage{graphics}
\usepackage{graphicx}
\usepackage{subfigure}
\usepackage{epsfig}
\usepackage{amssymb}
\usepackage{amsthm}

\journal{Photonics and Nanostructures}

\begin{document}

\begin{frontmatter}

\title{On effective electromagnetic parameters of artificial nanostructured magnetic materials}

\author{C.R. Simovski and S.A. Tretyakov}

\address{Department of Radio Science and Engineering
/ SMARAD Center of Excellence, Aalto University, School of Science
and Technology, 00076, Aalto, Finland}

\begin{abstract}
In this paper we discuss effective material parameter description of
new nanostructures designed to perform as artificial magnetic
materials for visible light. Among these structures there are
various split-ring resonators, dual-bar structures, fishnet layers
and other geometries. Artificial magnetic response in these
structures appears due to weak spatial dispersion effects, and it is
important to study the conditions under which the magnetic response
can be adequately measured with effective permeability tensor. On
the examples of dual bars and split rings we show that this is
possible only under some quite restrictive conditions. In the
general case, more complicated constitutive relations with more
effective material parameters need to be developed.

\end{abstract}


\end{frontmatter}

\def\e{\begin{equation}}
\def\f{\end{equation}}
\def\%#1{\mbox{\boldmath $#1$}}
\def\=#1{\overline{\overline #1}}
\def\*#1{\overline{\overline{\overline #1}}}
\def\s{\strut\displaystyle}
\def\_#1{{\bf #1}}
\def\o{\omega}
\def\va{\varepsilon}
\def\M{\mu}
\def\d{\nabla}
\def\p{\partial}
\def\rot{\nabla\times}
\def\div{\nabla\cdot}
\def\.{\cdot}
\def\x{\times}
\def\##1{{\bf#1\mit}}
\def\Re{{\rm Re\mit}}
\def\Im{{\rm Im\mit}}
\def\l#1{\label{eq:#1}}
\def\r#1{(\ref{eq:#1})}
\def\am{\left(\begin{array}{c}}
\def\amm{\left(\begin{array}{cc}}
\def\a{\end{array}\right)}
\def\ds{\displaystyle}


\section{Introduction}

Conventionally, the electromagnetic properties of bulk materials are
described by their permittivity ($\varepsilon$) and permeability
($\mu$) tensors. There are no means for the direct measurement of
these electromagnetic parameters, which are for this reason
dependent on the model used for their extraction from the
measurement data. Approaches to measurements of even these
fundamental electromagnetic characteristics of materials vary
dramatically for electromagnetic waves of different frequencies,
e.g. radio waves and light, and require specialized measurement
techniques. These difficulties are dramatically escalated in
electromagnetic characterization and metrology of nanostructured
materials with complex geometries of unit cells and exotic
electromagnetic response. For example, let us consider a device that
operates at the wavelength of 500-600 nm, with the structural
periodicity of artificial material in the range of  50-100 nm, while
the sizes of resonant inclusions are about 30-70 nm. Such
structures, if they are bulk (3D) lattices, can still be described
in terms of effective parameters (permittivity and permeability) of
an equivalent homogeneous medium. However, the conventional models
based on quasi-static homogenization procedures normally applied at
the atomic level are not applicable here, and the resulting
effective phenomenological parameters (like permittivity and
permeability) often have quite different physical meaning as
compared with conventional materials -- natural ones and composites
of non-resonant inclusions.

The bulk material parameters $\varepsilon$ and $\mu$ of an effective
medium (isotropic or anisotropic) are classically defined so that
they relate vectors $\_D$ and $\_B$ of the electromagnetic field in
a medium with macroscopic (averaged) vectors $\langle\_E\rangle$ and
$\langle\_H\rangle$, respectively: \e
\_D=\va_0\langle\_E\rangle+\_P\equiv \va_0\va\cdot
\langle\_E\rangle,\quad \_B=\mu_0(\langle\_H\rangle+\_M)\equiv
\mu_0\mu \cdot \langle\_H\rangle \l{defs}\f Here $\_P$ and $\_M$ are
the bulk electric and magnetic polarizations resulting from
microscopic electric and magnetic dipole moments of particles
averaged over the same volume as the electric and magnetic fields.
For regular arrays of atoms, molecules or inclusions this is usually
the volume $V$  of the lattice unit cell. Parameters defined by
formulas \r{defs} are applicable for solving boundary problems with
the Maxwell boundary conditions \cite{Born,Jackson,agran}.

Physically sound local material parameters should satisfy the
following basic requirements:

\begin{itemize}
\item Passivity (for the temporal dependence $e^{-i\omega t}$ it
implies ${\rm Im}(\varepsilon)>0$ and ${\rm Im}(\mu)>0$
simultaneously at all frequencies; for the $e^{j\omega t}$ time
dependence convention the sign of both ${\rm Im}(\varepsilon)$ and
${\rm Im}(\mu)$ should be negative). Violation of this condition in
passive media (no sources of electromagnetic energy at frequency
$\omega$) means the violation of the second law of thermodynamics;
\item Causality (for media with negligible losses it corresponds to
conditions $\partial \left(\omega\varepsilon\right)/\partial
\omega>1$ and $\partial \left(\omega \mu)\right)/\partial \omega>1$.
This also means that in the frequency regions where losses are
small, material parameters grow versus frequency: $\partial
\left({\rm Re}(\varepsilon)\right)/\partial \omega>0$ and $\partial
\left({\rm Re}(\mu)\right)/\partial \omega>0$ );
\item Absence of radiation losses in electrically dense arrays with uniform
distribution of particles. This means that in lossless arrays the
electromagnetic parameters should take real values.
\end{itemize}
The first two requirements (passivity and causality) are most known
and can be mathematically expressed through the Kramers-Kronig
relations (see e.g. in \cite{Born,Jackson,Landau}). These basic
physical requirements must be satisfied to ensure that the use of
the effective medium description does not lead to nonphysical
results. Furthermore, locality of the model implies that the
parameters
\begin{itemize}
\item are independent of the spatial distribution of fields excited in
the material sample,
\item  are independent of the geometrical size and shape of the
sample.
\end{itemize}
This ensures independence of material parameters on the wave vector
$\_q$, if one uses a plane wave as a probe field to determine the
material parameters. For a given frequency this means independence
of effective parameters on the wave propagation direction
\cite{Landau}.  These two requirements are difficult to satisfy for
many nanostructures, in part because the samples usually contain
only a few (or even one) layers of inclusions or patterned surfaces
across the sample thickness. This introduces limitations on the
applicability area of the effective parameters and demands the use
of alternative descriptions in terms of surface impedance or grid
impedance. If the last two conditions are not satisfied, the
material parameters can be used basically only for the same
excitation environment as in the characterization experiment.

Recently, considerable efforts have been devoted to the design and
realization of nanostructures behaving as artificial magnetic
materials at terahertz and optical frequencies, with the main
motivation to create double-negative materials (e.g.,
\cite{fishnet1,fishnet11,fishnet2}). These structures are usually
characterized using a probe plane electromagnetic wave normally
incident at planar samples. Then, the classical Nicolson-Ross-Weir
method \cite{NRW1,NRW2} is used to extract the effective
permittivity and permeability of the sample. However, in many
situations one can observe that the resulting parameters do not
satisfy the basic physical requirements of local material parameters
(parameters have nonphysical signs of the imaginary parts in some
frequency regions, do not satisfy the Foster theorem (e.g.,
\cite{Collin}) in low-loss regions, depend on the incidence angle of
the probing wave, etc.) In a recent paper \cite{Menzel} is was shown
that for typical optical fishnet structures the effective medium
description in terms of permeability is not valid in the whole range
where the backward-wave regime is observed. While these parameters
still correctly restore the reflection and transmission coefficients
for plane waves which were used in the characterization, they fail
to describe the material for other excitations and sample shapes.
This dramatically reduces their value in the design of applications
which would utilize the effective magnetic properties of the new
materials. Thus, it is of scientific and practical importance to
understand the limitations of the effective parameter models and
study the reasons for nonphysical behavior of extracted parameters.
In this paper we make a step in this direction analyzing magnetic
response of some simple inclusion geometries, conventionally used in
the design of artificial magnetics.

\section{Effective medium description of bulk optically dense metamaterials}

In this paper we concentrate on effective-parameter characterization
of optically dense bulk nanocomposites formed by electrically small
scatterers. Even in this case the homogenization is a difficult task
that obviously implies answering to the following questions:
\begin{itemize}
\item How to introduce (define) material parameters of composite
media, what is the physical meaning of them and in which
electrodynamic problems such material parameters are applicable?
\item What are frequency bounds in which these material parameters
keep their physical meaning and applicability? \item What are the
physical limitations that should be imposed on these material
parameters and can be used as a check of calculations, measurements,
and, finally, in practical applications?
\end{itemize}

Frequency dispersion in metamaterials formed by electrically
(optically) small inclusions can be strong when particles are
excited in a vicinity of their resonant frequencies. This implies
dramatic shortening of the wavelength inside the composite, which
means that spatial dispersion effects can appear even in optically
dense metamaterials. If the effective wavelength $\lambda/n_{\rm
eff}$, where $n_{\rm eff}$ is the effective refraction index, is
close to or smaller than the lattice period, the local material
parameters cannot be introduced. This situation can be easily
detected experimentally or numerically, because in this case the
nonlocality in response is visible also in the properties of the
effective refraction index $n_{\rm eff}$ and wave impedance $Z_{\rm
eff}$. In the following, we assume that the lattice period is
electrically small, and concentrate on the analysis of the influence
of the inclusion shape on the effective properties.

\subsection{Artificial magnetism in composites of nanodimers}

It has been known since early fifties \cite{Schelk} that composites
containing electrically small conductive inclusions of complex
shapes can exhibit properties of artificial magnetics, usually
described by magnetic polarizability of inclusions $\alpha_{mm}$.
This parameter is defined as \e \_m=\alpha_{mm}\cdot \_H\f where
$\_m$ is the induced magnetic moment and $\_H$ is the local magnetic
field at the center of the particle. After appropriate averaging,
magnetic response of individual inclusions defines the effective
permeability of the composite. This magnetic response is a
manifestation of spatial dispersion effects, because the response to
magnetic field basically means the response to the nonuniform
component of the electric field (in form of $\nabla\x\_E$)
\cite{Kluwer,modeboo}. The key pre-requisite for the validity of the
effective permeability model is that the response to other
combinations of spatial derivatives of $\_E$ can be neglected
\cite{Serd,biama} (see also Chapter 2-1 of \cite{Filippo1}). Also,
electric quadrupole as well as other higher-order polarization
moments should be negligibly small as compared with the magnetic
dipole \cite{biama}. This very much depends of the inclusion shape.
In fact, only for some specific shapes these conditions can be
satisfied.

\begin{figure}
\begin{center}
\includegraphics[width=125mm]{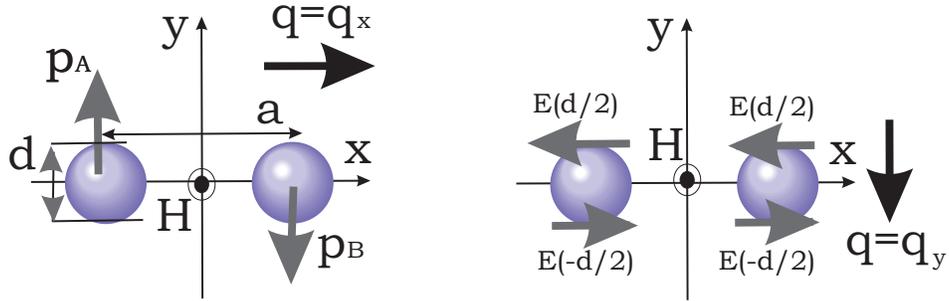}
\caption{(Color online) A unit cell of a lattice of pairs of
plasmonic spheres. If the wave propagates along $x$, the magnetic
field induces a resonant magnetic moment (and an electric
quadrupole moment). If the wave propagates along $y$, the same
magnetic field at the same frequency induces only an octupole
moment, which is negligible for optically dense lattices. Such a
lattice cannot be described only in terms of $\varepsilon$ and
$\mu$.} \label{figure}
\end{center}
\end{figure}

Let us, for example, consider arrays of dual plasmonic nanoparticles
(dimers): dual spheres, dual bars or other nanostructures in which
the resonant magnetic response of a unit cell is related with the
phase shift of the exciting wave over the distance $a$ between two
plasmonic elements. In one of the resonant modes of such pairs the
electric polarizations induced in two particles are out of phase,
which corresponds to high magnetic moment of the pair. Usually,
composites formed by many such inclusions are modelled by effective
permeability and permittivity, as in
\cite{dual0}--\cite{dual7} and a number of other papers. However, we
will see from the following that in this case the physical meaning
of the permeability introduced via averaging of the induced magnetic
moments is different from the conventional definition, which limits
the possible use of this effective parameter.

Fig.~\ref{figure} represents a plasmonic nanodimer. Let this
nanopair be a unit cell of a nanostructured metamaterial. The
magnetic moment of the medium unit cell of volume $V$ is defined as
\cite{Jackson,Landau} \e \_m={1\over 2}\int_V \_J\x {\_r} \, dV
\l{Mag}\f Here $\_r$ is the radius-vector referred to the particle
center, and $\_J=
-i\omega\varepsilon_0(\varepsilon-\varepsilon_h)\_E$ is the
polarization current density (in our case that inside the plasmonic
spheres). Metal nanospheres are non-magnetic, and their permittivity
is denoted as $\varepsilon$. $\varepsilon_h$ is the permittivity of
the host medium. The response of the pair of nanopsheres to the
time-varying magnetic field is in fact the response of the nanopair
to spatially varying electric field \cite{Kluwer,modeboo}. When the
local electric field $\_E$ acting on different nanospheres is
different, the dimer can acquire magnetic moment. Since for the
centers of two nanospheres the radius vector is equal to $\_r=\pm
(a/2)\_x_0$, magnetic moment $m$ of the nanopair given by \r{Mag} is
always $z$-directed and for any direction of the wave propagation
equals to $ m=m_z=-(i\omega a/2)[p_y(B)-p_y(A)]$. The action of the
local magnetic field $\_H$ to the nanopair is in this way
represented as action of electric fields $\_E(A)$ and $\_E(B)$ to
nanospheres A and B.

To quantify the magnetic effect, we assume that the dimer is probed
by a pair of plane waves that form a standing wave. This allows us
to position the dimer center at the point where the incident
electric  field is zero and this way find the response to
quasi-uniform magnetic field \cite{Kluwer,modeboo}. If the exciting
waves propagate along $x$ ($q=q_x$ in Fig.~\ref{figure}), the local
electric field is $y$-polarized, $E(B)=-E(A)$, and $p(B)=-p(A)$.
Respectively, $m=-i\omega ap(B)/2$. The local magnetic field $H=H_z$
at the nanopair center is related to the electric fields $E(A)$ and
$E(B)$ through Maxwell's equations: \e i\omega \mu_0
H_z=\frac{\partial E_y}{\partial x}-\frac{\partial E_x}{\partial
y}=\frac{\partial E_y}{\partial x}\approx
\frac{E(B)-E(A)}{a}=\frac{2E(B)}{a} \l{iden}\f In \r{iden} we have
taken into account the small optical size of the nanopair
($|k|a<1$). The nanopair magnetic polarizability
$\alpha_{mm}=\alpha_{mm}^{zz}$ is then equal to
\begin{equation}
\alpha_{mm}\equiv \frac{m}{H}=Z_0(ka)^2\frac{p_B}{E(B)}=Z_0(ka)^2
\alpha \l{first}
\end{equation}
where $Z_0$ is the free-space wave impedance and $\alpha =
E(B)/p(B)=E(A)/p(A)$ is the electric dipole polarizability of the
individual plasmonic particle. The magnetic polarizability in this
special case of propagation turns out to be resonant at the same
frequency as the plasmon resonance of an individual particle.
Actually, the magnetic resonance is slightly red-shifted with
respect to that of the single sphere (our simplistic model does not
take into account the mutual coupling of spheres), but it is not
important for our purposes.

On the other hand, if the probing waves propagate along $y$ ($q=q_y$
in Fig.~\ref{figure}), the response of the nanopair to \emph{the
same} magnetic field $\_H$ at the dimer center corresponds to zero
induced magnetic moment. In this case we have \e i\omega \mu_0
H_z=\frac{\partial E_y}{\partial x}-\frac{\partial E_x}{\partial
y}=-\frac{\partial E_x}{\partial y} \l{iden1}\f The magnetic moment
of the unit cell induced by the magnetic field is zero because $E_x$
is zero at the centers of nanopsheres and their electric dipole
moments vanish. Magnetic field $\_H$ induces opposite electric
polarizations of the upper and lower halves of both spheres as it is
shown in Fig. \ref{figure}. This effect is practically negligible
and has nothing to do with magnetic polarization, i.e., in this case
$\alpha_{mm}=0$.

We observe a dramatic dependence of the magnetic polarizability
$\alpha_{mm}$ on the direction along which the incident field is
changing. For example, in paper \cite{dual7} this effect is treated
as that of strong spatial dispersion, resulting in both permittivity
and permeability strongly dependent on the propagation direction.
However, the optical size of the dimers and the lattice period are
assumed to be optically very small, and there are no physical
reasons for strong spatial dispersion in the effective medium
response.
The absence of strong spatial dispersion means that the permeability
tensor $\mu$ (as well as $\va$) cannot depend strongly on the
direction of the vector $\_q$.

The answer to this paradox is given in the theory of multipole
media, also called media with weak spatial dispersion
\cite{Serd,biama,Filippo1,multipole}. Multipole media depend on
higher-order multipole response accompanying the magnetic dipole
response of complex particles. In the present case this higher-order
multipole response is the electric quadrupole susceptibility. In
\cite{dual7} it was correctly noticed that the quadrupole
polarization of plasmonic dimers is as essential as their electric
and magnetic dipole polarizations.
In general, media with resonant quadrupole moments as any other
multipole media cannot be described only in terms of $\varepsilon$
and $\mu$. Physically sound material equations for multipole media
contain besides macroscopic fields $\_E$ and $\_H$ also spatial
derivatives of $\_E$ \cite{Serd,biama,multipole}. Respectively, more
material parameters are required. In this theory the unit cell
magnetization can be properly described without involving spatially
dispersive permittivity $\varepsilon(\_q)$ and permeability
$\mu(\_q)$, which describe only excitation by plane waves traveling
along one particular direction.

For a medium formed by dimers we can write for any direction of the
wave propagation \e m=m_z=-i\omega a{p_y(B)-p_y(A)\over 2}={-i\omega
a^2\alpha\over 2} {E_y(B)-E_y(A)\over a}\approx 2\nu \nabla_x E_y
\l{spec}\f where $\nu= i\omega a^2\alpha/4$. This relation is more
adequate than the proportionality between $m$ and $H$, which is only
a special case of \r{spec}. We can  present \r{spec} as the sum of
the symmetric and antisymmetric derivative forms: \e
m_z=\nu(\nabla_x E_y-\nabla_y E_x)+\nu(\nabla_x E_y+\nabla_y E_x) \f
or, in the index form, \e
m_{\alpha}=\Gamma_{\alpha\beta}H_{\beta}+\kappa_{\alpha\beta\gamma}\nabla_{\beta}E_{\gamma}
\l{eq}\f In \r{eq} indices $\alpha,\ \beta,\ \gamma$ correspond to
the Cartesian coordinates and we use the notations \e
\Gamma_{zz}={-i\nu\omega\mu_0},\quad \kappa_{zxy}=\kappa_{zyx}=\nu
\l{res}\f Components with all the other combinations of indices
$\alpha,\ \beta,\ \gamma$ equal zero. From \r{eq} it is clear that
the magnetic polarization contains two parts. The first part is the
magnetic moment of the dimer induced by the magnetic field which
results in the resonant effective permeability of the composite
medium. The second part describes the magnetic moment of the dimer
which is proportional to all the other combinations of spatial
derivatives of the electric field. When the wave propagates along
$y$, the second term cancels out with the first one and the induced
magnetic moment vanishes. The dependence of the magnetic
\emph{moment} on the propagation direction is not a feature of
strong spatial dispersion because the \emph{polarizability} tensors
$\Gamma_{\alpha\beta}$ and $\kappa_{\alpha\beta\gamma}$ do not
depend on the propagation direction. This feature of weak spatial
dispersion was discussed in books \cite{Serd,biama}. The material
equation for the magnetic field resulting from \r{eq} and the
quasi-static averaging procedure can be written in the index form as
 \e B_{\alpha}\equiv \mu_0 H_{\alpha}+M_{\alpha}= \mu_0\mu_{\alpha\beta}H_{\beta}
 +\chi_{\alpha\beta\gamma}\nabla_{\beta}E_{\gamma}
 \l{mat1}\f
Here tensor $\mu_{\alpha\beta}$ results from $\Gamma_{\alpha\beta}$
and tensor $\chi_{\alpha\beta\gamma}$ results from
$\kappa_{\alpha\beta\gamma}$.

A similar consideration can be done for the quadrupole moment of the
dimer. The vectors of the electric dipole and magnetic dipole
moments of the dimer and the tensor of its quadrupole moment are
resonant. This means that the electric displacement vector $\_D$
essentially includes the contribution of the electric quadrupole
polarization $Q_{\alpha\beta}$. This evolves one more material
parameter in the material equation which takes the form
\cite{Serd,biama,Filippo1,multipole}:
 \e D_{\alpha}\equiv \va_0 E_{\alpha}+P_{\alpha}+{1\over
 2}\nabla_{\beta}Q_{\alpha\beta}=\va_0\va_{\alpha\beta}E_{\beta}+\xi_{\alpha\beta\gamma\delta}\nabla_{\beta}
 \nabla_{\gamma}E_{\delta}
 \l{mat2}\f
Material equations \r{mat1} and \r{mat2} corresponds not only to
dimers. They refer to all composites whose inclusions possess
resonant quadrupole polarization in the absence of bianisotropy, see
e.g. \cite{S1,S2,S3}.

In this formalism the boundary conditions should be revised as
compared to the formalism of initial equations \r{mat1} and
\r{mat2}, because  the set of Maxwell's boundary conditions is not
enough to solve boundary problems for such media. Additional
boundary conditions should be derived as it was explained in books
\cite{Serd,biama}. However, to our knowledge, for media described by
equations \r{mat1} and \r{mat2}  boundary conditions have not been
derived. In the known papers composites of dimers and other
multipole media are described by only $\varepsilon$ and $\mu$. That
description implies the use of Maxwell's boundary conditions, but
can be applied only to the plane-wave incidence at the angle used in
the extraction of the effective parameters.

\subsection{Artificial magnetism in bianisotropic composites}

Next we will discuss the artificial magnetism in composites based on
split rings (SRR), with the emphasis on the effect of bianisotropy.
In particular, we consider an array of U-shaped metal SRRs
experiencing plasmonic resonance in the optical range. Similarly to
the previous section, we use a pair of plane waves to probe the
particle response to external magnetic fields. First, let us show
that magnetic polarizabilities are different for the two cases: when
the waves propagate with vector ${\_q}=q\_x_0$ along $x$ and the
electric field associated to the local magnetic field $\_H$ is
$y$-polarized, and when $\_q={q}\_y_0$ and the same magnetic field
is related to $x$-polarized electric fields.

The following consideration is based on the method of induced
electromotive force (IEF) as it was formulated  in \cite{Schelk1}
for conducting wire scatterers. A formula for the IEF more
appropriate to the case of a plasmonic SRR is more involved and
corresponding derivations are cumbersome, whereas the result will be
definitely similar to that obtained below.

\begin{figure}
\begin{center}
\includegraphics[width=0.7\textwidth]{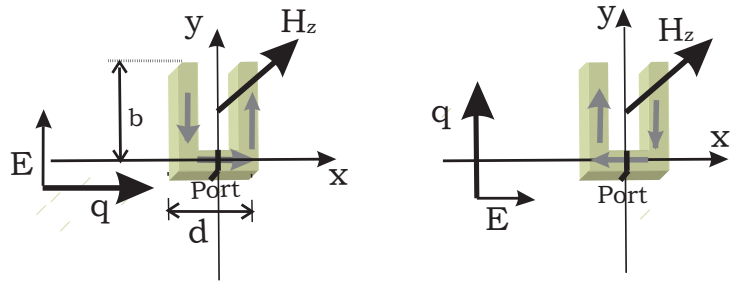}
\caption{(Color online) A sketch of U-shaped SRRs for two cases of
the wave propagation. The incident magnetic field at the particle
center is the same in both cases. Introducing an imaginary port at
the center of the horizontal bar and applying the method of
induced electromotive force we prove the strong difference of the
particle magnetic polarizabilities on the propagation direction
for these two cases.} \label{figure1}
\end{center}
\end{figure}

Let us choose the coordinate system as shown in Fig.~\ref{figure1}
and locate an imaginary port in the origin. The electromotive force
induced by the local electromagnetic field at this imaginary port is
equal to \cite{Schelk,Schelk1} \e {\cal E}={1\over I_0}\int_l E_l
I_{\rm rad}(l)\ dl  \l{emf1}\f Here $E_l$ is the component of the
local electric field tangential to the scatterer's contour, $I_{\rm
rad}$ is the complex amplitude of the linear current distributed
over this contour in the radiation regime, when an external voltage
source is connected to the port, and $I_0=I_{\rm rad}(0)$. In the
case $q=q_x$ from Maxwell's equation \r{iden} we obtain \e E_l=\pm
\left. E_y\right|_{x=\pm {d\over 2}}=\pm i{\omega\mu_0 Hd\over 2}
\l{el1}\f Since the U-shaped scatterer is optically small and the
currents at the ends of the stems vanish, the distribution of the
induced current over the scatterer's contour $I_{\rm rad}(l)$ can be
approximated as a linear function of the contour coordinate $l$: \e
I_{\rm rad}(l)=I_0{b+{d\over 2}-l\over b+{d\over 2}} \l{cur}\f where
$l=0$ at the port, $l=|x|$ for $-d/2<x<d/2,\ y=0$ and $l=y+d/2$ for
$x=\pm d/2,\ y<b$. Upon substitution of \r{el1} and \r{cur} formula
\r{emf1} can be rewritten in the case $q=q_x$ as \e {\cal
E}=i{\o\mu_0 Hd\over b+{d\over 2}}\int\limits_0^b (b-y)\
dy=i{\o\mu_0 Hb^2d\over 2b+d} \l{emf2}\f The same electromotive
force $\cal E$ can be expressed through the input impedance $Z_{\rm
in}$ of the scatterer referred to the imaginary port: ${\cal E}=I_0
Z_{\rm in}$. Equating these two representations  one can express the
ratio $(I_0/H)$ through $Z_{\rm in}$: \e {I_0\over H}= i{\o\mu_0
Hb^2d\over Z_{\rm in}(2b+d)}\l{zin1}\f

When we calculate $\alpha_{mm}$, we use the definition \r{Mag}, and
take into account that the current distribution induced by the local
field is different from the distribution of the current excited by
an external voltage in the port. Therefore, instead of \r{cur} we
use a more adequate approximation which is smooth at the coordinate
origin: \e I(l)\equiv I_0F(l)=I_0{\left(b+{d\over
2}\right)^2-l^2\over \left(b+{d\over 2}\right)^2}\l{ade}\f Using the
definition of the magnetic moment \r{Mag} and the distribution
function $F(l)$ we can write the magnetic polarizability of the SRR
in the form: \e \alpha_{mm}= {m\over H}={\mu_0 I_0\over
2H}\left[{b\over 2}\int\limits_{-d/2}^{d/2}F(x,y=0)\ dx +2{d\over 2}
\int\limits_{0}^{b}F\left(x={d\over 2},y\right)\ dy \right]
\l{amm1}\f  Substituting relation \r{ade} for function $F(l)$ into
\r{amm1} we finally obtain \e \alpha_{mm}= i{\omega\mu_0^2
b^3d^2(6b+d)\over 2Z_{\rm in}(2b+d)^2} \l{mmm} \f

In the case $q=q_y$ from Maxwell's equation \r{iden1} we find  \e
E_l=E_x(y=0)=\pm i{\o\mu_0 Hb\over 2}\l{el2}\f Substituting \r{el2}
and \r{cur} into the main formula \r{emf1} we obtain \e {\cal
E}=i{\omega\mu_0 Hb\over b+{d\over
2}}\int\limits_0^{d/2}\left({d\over 2}+b-x\right)\ dx=i{\omega\mu_0
Hbd(4b+d)\over 8(2b+d)} \l{emf3}\f Only the horizontal bar gives
contribution into the induced electromotive force, while the
integrals over the stems vanish. For the ratio  $I_0/H$ we have in
this case \e {I_0\over H}= i{\o\mu_0 bd(4b+d)\over 8Z_{\rm
in}(2b+d)}\l{zin2}\f Substituting \r{zin2} into formula \r{amm1} we
obtain \e \alpha_{mm}= i{\omega\mu_0^2 b^2d^2(6b+d)(4b+d)\over
16Z_{\rm in}(2b+d)^2} \l{mmm1} \f

This result clearly differs from \r{mmm}. The ratio of magnetic
polarizabilities corresponding to the two cases of propagation is as
follows: \e {\alpha_{mm}^{(q=q_x)}\over
\alpha_{mm}^{(q=q_y)}}={8b\over 4b+d}\l{rat}\f For $b=d$ these
magnetic polarizabilities differ by the factor of $1.6$. Though in
both cases of the wave propagation direction the magnetic
polarizability is nonzero, this example clearly shows that the
magnetic polarizability $\alpha_{mm}$ of a U-shaped SRR is an
ambiguous value, similarly to the case of a dimer. Respectively, one
should take into account the electric quadrupole polarization.

\subsection{Bianisotropic material relations for nanostructured metamaterials}

When some local magnetic field acts on a U-shaped SRR, the same
polarization current which forms dipole moments of the stems (whose
vector sum vanishes) also creates some  noncompensated resonant
dipole moment of the horizontal bar. Thus, local magnetic field in
this case creates resonant magnetic dipole moment, resonant electric
quadrupole moment, and resonant electric dipole moment. The effect
of electric dipole polarization of the medium unit volume by
magnetic field can be expressed as the second term in the right-hand
side of the equation \e
P_{\alpha}=\Lambda_{\alpha\beta}E_{\beta}+\Psi_{\alpha\beta}H_{\beta}
\l{fi1}\f This effect in reciprocal media is complemented by the
effect of the magnetic polarization of the medium by electric
fields: \e
M_{\alpha}=\Theta_{\alpha\beta}H_{\beta}+\Psi'_{\alpha\beta}E_{\beta}
+\kappa_{\alpha\beta\gamma}\nabla_{\beta}E_{\gamma}\l{fi2}\f This
couple of effects (tensors $\Psi$ and $\Psi'$ are uniquely related:
$\Psi'_{\alpha\beta}=-\Psi_{\beta\alpha}$) is well known in the
electromagnetic theory and engineering (e.g., \cite{Serd}) and is
called bianisotropy. Media formed by parallel U-SRR particles
possess resonant bianisotropy. Even dual SRRs introduced in
\cite{SRR} are bianisotropic, and their bianisotropy was widely
discussed in the literature after the publication of paper
\cite{Ric}. The bianisotropy of dual SRRs is an important factor
that can suppress the backward-wave regime in metamaterials
\cite{Ric}. Moreover, U-shaped particles show stronger bianisotropic
effects than dual SRRs. Since bianisotropy is a stronger effect than
artificial magnetism (first- and second order effects in terms of
$ka$), characterization of such metamaterials  only by two
parameters $\varepsilon$ and $\mu$ is clearly not adequate.

From \r{fi1} and \r{fi2} it is clear that two additional material
parameters should be introduced compared to multipole media without
bianisotropy. Material equations \r{mat1}, \r{mat2} for
bianisotropic multipole media generalize to the form:
 \e
 B_{\alpha}=\mu_0\mu_{\alpha\beta}H_{\beta}+\psi_{\alpha\beta}E_{\beta}+
\chi_{\alpha\beta\gamma}\nabla_{\beta}E_{\gamma}\l{mat111}\f
 \e D_{\alpha}=\va_0\va_{\alpha\beta}E_{\beta}
-\psi_{\beta\alpha}E_{\beta}+
 \xi_{\alpha\beta\gamma\delta}\nabla_{\beta}
 \nabla_{\gamma}E_{\delta}
 \l{mat222}\f
The material parameter describing the bianisotropic effect in
equations \r{mat111} and \r{mat222} is tensor $\psi$, called
\emph{magnetoelectric coupling} parameter in the theory of
reciprocal bianisotropic media (e.g. in \cite{Serd,biama,biaha}).

Two special cases of bianisotropic media are well known: chiral
media when tensor $\psi$ is symmetric, and omega media when tensor
$\psi$ is antisymmetric. According to the classification
\cite{O,Serd} arrays of parallel U-SRRs belong to the class of omega
media. The known theory of omega media (see e.g. in
\cite{Serd,O1,O2,O3}) was developed for the special case when
tensors $\chi$ and $ \xi$ in \r{mat111} and \r{mat222} are
negligible. Fabrication of nanostructured arrays of U-SRRs \cite{U}
will hopefully motivate researchers to develop the theory of
multipole bianisotropic media, and the first task should be the
derivation of boundary conditions for media described by equations
\r{mat1} and \r{mat2} and for media obeying equations \r{mat111} and
\r{mat222}.

\subsection{Artificial magnetism without resonant multipole
polarization}

Finally, let us discuss which inclusion shapes can potentially offer
artificial magnetism property without strong excitation of higher
multipoles. If the ring shown in Fig.~\ref{figure1} were closed, the
electric dipole moments in each side would be in all cases of the
wave propagation cancelled by that in the opposite arm. In an
electrically small closed loop with uniform current distribution,
the induced electromotive force is the same for any direction of
propagation of incident waves. This induced electromotive force is
equal to the negative time derivative of the applied magnetic flux.
No bianisotropy and no quadrupole moment arise, and the array of
closed rings (at enough low frequencies, of course) behaves an
effective diamagnetic medium.

If the open part of the ring is very small compared to its
perimeter, the dependence of the induced magnetic moment on the
propagation direction is also small and, respectively, quadrupole
moments are negligible. However, the bianisotropy remains strong and
resonant, since the induced current distribution is not uniform,
which results in both electric and magnetic dipole moments of the
loop. The bianisotropy of the array related to the open part of the
loop can be suppressed in two ways. The first way is to prepare the
unit cell from at least two SRRs so that the bianisotropy of the
unit cell is compensated. That way was suggested in \cite{Ric}. A
similar approach was developed in \cite{Omega1,Omega2} where a
practically isotropic doubly-negative composite was designed. The
unit cell in these works was formed by six properly arranged
$\Omega$-shaped metal particles. A dense stack of split rings with
intermitting positions of splits was introduced in
\cite{metasolenoid}. In all these design the governing idea is to
realize uniform distribution of induced current along the inclusion
perimeter.

However, the above designs refer to the microwave range. Realistic
approaches to the artificial magnetism without strong spatial
dispersion, high-order multipoles and bianisotropy in the optical
range were suggested more recently. Suitable design solutions are,
for example, as follows: four-split optical SRRs suggested in
\cite{4cutSRRs} and effective nanorings of plasmonic spheres
\cite{Engheta2006}. In \cite{Engheta2006} an effective ring is
formed by plasmonic nanospheres located at the corners of a regular
polygon. The minimal number of nanospheres when the higher
mutlipoles can be (with a qualitative accuracy) neglected is four.
In \cite{4cutSRRs} each of four plasmonic scatterers forming the
effective ring is a planar $L$-shaped nanoparticle. In
\cite{Engheta2006} these four plasmonic scatterers are spheres. It
is clear that in both these design solutions the physical mechanism
of the artificial magnetism is the same.

Notice, that the concentration of these effective rings should be
very high in order to obtain a strong resonance of permeability.
Results presented in \cite{Engheta2006} correspond to almost
touching nanorings. Therefore the magnetism of such nanorings can be
realized only in the anisotropic variant. The requirement of the
high concentration makes a composite of randomly oriented effective
rings of nanospheres hardly feasible (the same concerns three
mutually orthogonal arrays of effective rings). The fully isotropic
solution based on core-shell magnetic nanoclusters
\cite{Simovski2009} is probably more practical.

\section{Conclusion}

In this paper we have considered the problem of effective medium
description of nanostructures designed to emulate properties of
magnetic materials in the visible range. The artificial magnetic
effect can be achieved due to specific shapes of nanoinclusions and
results from weak spatial dispersion in the composite. In general,
spatial dispersion produces a variety of effects, including resonant
bianisotropy, and the artificial magnetism is only one of them. Care
should be taken in introducing effective permeability in such a way
that it indeed measures the averaged magnetic polarization and can
be used in solving boundary problems for composite samples. If the
inclusion geometry is such that bianisotropic effects are allowed,
appropriate terms should be added to the material relations:
bianisotropic effects are usually stronger than the artificial
magnetism. Except some very specific geometries, induced electric
quadrupoles are of the same order as magnetic moments and they must
be included in the analysis. Furthermore, not only the response to
the curl of electric field (that is, to magnetic field), but also to
all other combinations of the first-order spatial derivatives of
electric field should be included for adequate description of the
effective material properties. In other words, characterizing bulk
optically dense nanostructured metamaterials starts from the
analysis of multipoles and bianisotropy. The insight on adequate
material equations for particular nanocomposite is mandatory for
physically sound characterization. Unfortunately, the effective
medium theory of weakly dispersive media is not complete at this
stage, and further investigations are necessary (especially in what
concerns boundary conditions at interfaces), before we will be able
to use the effective parameters in the design of optical devices.

\section*{Acknowledgement}

\noindent This work has been partially funded by the Academy of
Finland through the Center-of-Excellence program and supported by
the FP7 CSA \emph{ECONAM}.

\end{document}